\documentstyle[epsfig,amssymb,amsmath,array,multirow]{article}

\newcommand{\ra}{\rangle}

\title{RISQ - reduced instruction set quantum computers}
\author{Klaus M\o lmer and Anders S\o rensen\\
{\small Institute of Physics and Astronomy, University of Aarhus,
DK-8000 \AA rhus C, Denmark}}
\begin{document}
\maketitle

\begin{abstract}

\noindent
Candidates for quantum computing which offer only
restricted control, {\it e.g.}, due to lack of access to 
individual qubits, are not useful for {\it general purpose} 
quantum computing. We present concrete proposals
for the use of systems with such
limitations as RISQ - {\it reduced instruction set} 
quantum computers and devices - for simulation of quantum dynamics,
for multi-particle entanglement and squeezing of collective spin variables.
These tasks are useful in their own right, and they also  
provide experimental probes for the functioning of quantum gates 
in pre-mature proto-types of quantum computers.
\end{abstract}

\section{Introduction}

Notwithstanding dedicated theoretical and experimental efforts,
progress in practical implementation of quantum computing is not
advancing rapidly. 
Quantum computing is based on the superposition principle which,
when applied  to an information carrying system, suggests that
parallel processing of a large number of states, {\it e.g.},
to be identified with inputs to a function, is possible. 
Only recently it was realized
that unitary evolution of superposition states followed by
measurements, allowed by quantum theory, provides computational
powers, exceeding the one of classical computers \cite{shor,grover}.
Since then many strategies for practical quantum computing 
have been investigated.
Due to the coincidence of few-qubit quantum computing and 
ingredients of the advanced 
field of optimal control \cite{control}, already exercised extensively 
in molecular magnetic resonance spectroscopy, researchers in that field
have been able 
to rapidly implement a number of theoretical quantum computing proposals
\cite{nmr}, but the molecular systems are not promising for larger scale
computation.  
Ingredients from the past years' successful experimental control of 
atoms and single quantized field modes \cite{qfm,qfm2}, 
trapped ions \cite{ions}, donor spin states in solids \cite{kane},
quantum dots \cite{dots}
and Josephson junctions \cite{josephson} have been tailored to yield 
proposals for single qubit and two-qubit operations, scalable, 
in principle, to quantum computation with an arbitrary number of qubits. 

Experimental research groups now study these
proposals, and it is clear that we will see much
progress in the coming years, but also that quantum
computing is not going to be easy. 
The development of quantum computers and the progress 
of our research in quantum computing, is further hampered by the fact that
a small quantum computer is of little practical use, and so is
a large one which only ``gets the answer almost right" (in contrast to
one that ``gets the correct answer, but only sometimes"). 

We suggest
in this paper to identify applications of quantum computers
with reduced capabilities. We shall use the name RISQ computers
(Reduced Instruction Set Quantum computers) for such
devices, and we present examples of RISQ computers which may  be used to
solve quantum problems, much in the spirit of Feynman's
proposal for quantum computing \cite{feynman}. 
It was suggested by Lloyd \cite{lloyd} that the restriction of
physics problems due to locality and symmetries makes such a computer
potentially much less demanding to realize in practice than the general
purpose quantum computer. Indeed, we shall show that the reduced
capabilities of our RISQ systems may just coincide with these physical
restrictions, so that they do not present obstacles to ``Feynman quantum
computing".  We point out that 
for atoms and ions, entanglement produced by RISQ mechanisms
may improve spectroscopic resolution, atomic clocks, and length and frequency
standards. We believe that RISQ ideas will soon lead to 
operative and useful devices.

We shall focus on practical proposals. In Section 2, we present
a scheme for quantum gates and multi-particle entanglement
in ion traps,  which may be applied for 
full scale quantum computing, but which may also be applied in a RISQ version
without experimental access to individual ions in the trap. In Section 3,
we discuss quantum computing with neutral atoms in
optical lattices in a RISQ version without 
access to the individual atoms, useful for simulations of 
anti-ferromagnetism and for improvement of atomic clocks beyond the 
fundamental projection noise limit. Section 4 concludes the paper
with an optimistic view on quantum information processing as a tool in
physics.

\section{Ion trap quantum computers}

\subsection{General purpose quantum computing in ion traps}

At low
temperatures trapped ions freeze into a  crystal where 
the Coulomb
repulsion among the  ions  equilibrates the confining force from the
trapping potential. The vibrations of the ions are strongly  
coupled due to the Coulomb interaction, and in the harmonic oscillator
approximation they form a
set of collective vibrational modes.  One may excite one of these modes
by tuning a laser to one of the upper or lower sidebands of the
ions, {\it i.e.}, 
by choosing the frequency of the laser equal to the resonance frequency of
an internal transition in an ion plus or minus the vibrational frequency.
The laser is then on resonance with an  excitation of the internal
transition and a simultaneous change in  the vibrational
motion. This coupling of internal and 
external degrees of freedom has been extensively used for precise control
of the quantum state of trapped ions \cite{ions}, and in 1995 
Cirac and Zoller proposed that the ion trap can be used for quantum
computing \cite{cirac}. 

In the ion trap quantum computer
a qubit is represented by the internal states of an ion. Long
lived states, like for instance hyperfine structure states, are
preferred.  Single qubit rotation and two qubit gates are achieved by
focusing  a laser on each ion and by exploiting the collective vibrations for
interaction between the ions.  In the original proposal \cite{cirac} 
the system is restricted to the joint motional ground state of the ions. 
By tuning a laser to a sideband, a
vibration is  excited if the ion irradiated is in a certain 
internal state. Upon subsequent laser irradiation of
another ion,  the internal state of that ion is changed  only if 
the vibrational motion is excited. At the end of the resulting two-qubit
gate the vibrational
excitation is removed and additional gates may subsequently be implemented.

Under the assumptions of perfect access to the ions and 
complete absence of decoherence, the trapped ions can 
be used to compute any mathematical function, and since the 
ions can be initially set to a superposition of all  register states, one 
simultaneously obtains the evaluation of all function values - the
magic parallelism of quantum computing.
By electron shelving \cite{shelving}, the state of each qubit can be 
read out very effectively at the end of the calculation, if one can 
distinguish fluorescence from the different ions in the trap.

\subsection{Bichromatic excitation scheme}

We now describe our proposal \cite{bic1} for the efficient 
production of a two-qubit Hamiltonian like $\sigma_{y,i}\sigma_{y,j}$,
where the Pauli matrices acting on individual ions $i$ and $j$
represent the two-level qubit systems, {\it e.g.}, with $|0\rangle_i$
and $|1\rangle_i$ being the eigenstates of $\sigma_{z,i}$. A 
Hamiltonian proportional with $\sigma_{y,i}$ provides a rotation
between the two states $|0\rangle_i$ and $|1\rangle_i$,
and products of such operators provide conditional operations
which suffice to build a general purpose quantum computer.

\begin{figure}
\centerline{  \epsfig{angle=270,width=10cm,file=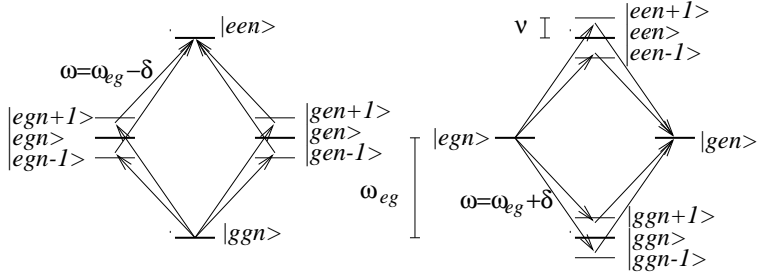}}
  \caption{Energy level diagram for two ions with quantized vibrational
   motion illuminated with bichromatic light.
    The one photon transitions indicated in the figure are not resonant,
   $\delta \neq \nu$, so only the two-photon transitions shown
from $|ggn\ra$ to $|een\ra$ and from $|egn\ra$ to $|gen\ra$
    are resonant.} 
  \label{detunings}
\end{figure}

We illuminate the two ions of interest with light of two
different frequencies, $\omega_{1,2}=\omega_{eg} \pm \delta$,
where $\omega_{eg}$ is the internal state transition frequency,
and $\delta$ is a detuning, not far from the trap frequency
$\nu$.
In Fig.~\ref{detunings}, we illustrate the action of such a 
bichromatic laser field on the state of the two ions of interest.
As shown in the figure,
the initial and final states $|ggn\rangle$ and $|een\rangle$, 
separated by $2\omega_{eg}=\omega_1+\omega_2$
are resonantly coupled, and so are the degenerate states $|egn\rangle$ and
$|gen\rangle$, where the first (second)  letter denotes the
internal state $e$ or $g$ of the $i^{th}$ ($j^{th}$) ion and $n$ is the 
quantum number for the relevant vibrational mode of the trap.
These resonant couplings lead to an effective 
Hamiltonian of the form
$\sigma^+_i\sigma^+_j+\sigma^-_i\sigma^-_j-
\sigma^+_i\sigma^-_j-\sigma^-_i\sigma^+_j \propto \sigma_{y,i}\sigma_{y,j}$.

\smallskip

\noindent
The restriction of the dynamics to the resonantly coupled states
applies in two interesting limits:

\subsubsection{Weak fields}

We can choose the fields so weak and 
the detuning from the sideband so large that the intermediate states
with $n\pm 1$ photons are not populated in the
process. 
It turns out \cite{bic1,bic3} that as long  as the ions 
remain in  the Lamb-Dicke regime, {\it i.e.},
their spatial excursions are  restricted to a small fraction of the
wavelength of the exciting radiation,  
the internal state transition is 
insensitive to the vibrational quantum number $n$.
This is due to interference between the interaction paths:
The transition via an upper sideband excitation $|n+1\rangle$,
has a strength of $n+1$ ($\sqrt{n+1}$ from
raising and  $\sqrt{n+1}$ from lowering the vibrational quantum
number), and the 
transition via $|n-1\rangle$ yields
a factor of $n$. 
Due to opposite signs of the intermediate state energy mismatch,
the terms interfere destructively, and the $n$
dependence disappears from the coupling. 

The coherent evolution of the internal atomic
state is thus insensitive to the vibrational quantum
numbers, and it may be observed with ions in any 
superposition or  mixture of 
vibrational states, even if the  ions  exchange vibrational 
energy with a surrounding reservoir. The control
of the thermal motion is of great difficulty in ion trap
experiments, and the tolerance to vibrations is a major asset
of our bichromatic proposal. 
In the RISQ section below, we show 
that the bichromatic gate can also be applied with interesting results 
without individual access to the ions in the trap, which removes
another technical complication for experiments.

\subsubsection{Strong fields}

In the Lamb-Dicke
limit with lasers detuned by $\pm\delta$ our bichromatic interaction
Hamiltonian can be written in the interaction picture with respect to
the atomic and vibrational Hamiltonian 
\begin{equation}
\label{interaction}
H_{\mbox{int}}=
-\sqrt{2}\eta\Omega J_y [x\cos(\nu-\delta)t +p\sin(\nu-\delta)t], 
\end{equation}
where we have introduced the dimensionless position 
and momentum operators for the centre-of-mass vibrational mode 
$x=\frac{1}{\sqrt{2}}(a+a^\dagger)$ and
$p=\frac{i}{\sqrt{2}}(a^\dagger-a)$, and where we have introduced
the collective internal state observable
$J_y = \frac{\hbar}{2} (\sigma_{y,i}+\sigma_{y,j})$ of the two ions
illuminated.
$\Omega$ is the Rabi frequency of the field-atom coupling, 
and $\eta$ is the Lamb-Dicke parameter.

The exact propagator
for the Hamiltonian~(\ref{interaction}) can be represented by the ansatz
\begin{equation}
  \label{u}
  U(t)={\rm e}^{-iA(t)J_y^2}{\rm e}^{-iF(t)J_y x}{\rm e}^{-iG(t)J_y p},
\end{equation}
where the Schr{\"o}dinger equation  $i\frac{d}{dt}U(t)=HU(t)$ leads to
the expressions $F(t)=-\sqrt{2}\eta\Omega\int_0^t \cos((\nu-\delta)t')dt', \ 
  G(t)=-\sqrt{2}\eta\Omega\int_0^t \sin((\nu-\delta)t')dt'$, 
and $A(t)=\sqrt{2}\eta\Omega\int_0^t F(t')\sin((\nu-\delta)t')dt'$.

If $F(t)$ and $G(t)$ both vanish after a period $\tau$, the propagator 
reduces to $U(\tau)={\rm e}^{-iA(\tau)J_y^2}$  at this instant,  {\it
  i.e.},  the vibrational motion is returned to its original
state, be it the ground 
state or any vibrationally excited state,  and we are left with an
internal state evolution which is {\it independent} of the external
vibrational state \cite{footnote}.
Note that $(\sigma_y)^2=1$ implies that $J_y^2 =
\frac{\hbar^2}{4}(2+2\sigma_{y,i}\sigma_{y,j})$, yielding precisely
the interaction that we need. 
The timing so that $G(\tau)$ and $F(\tau)$ vanish
allows faster gate operation than in Section 2.2.1, because we tolerate
that the internal state  is  strongly entangled 
with the vibrational motion in the course of the the gate. 

\begin{figure}[h]
\begin{center}
 \begin{minipage}{5cm}
  \epsfig{file=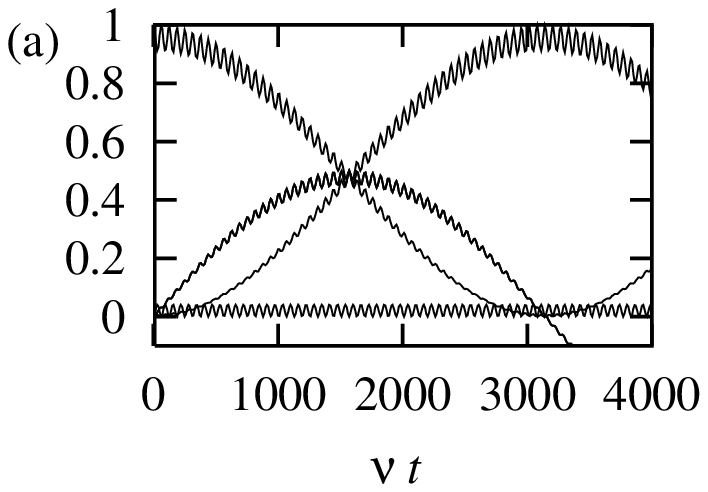,angle=270,width=5cm}
 \end{minipage}
 \begin{minipage}{5cm}
  \epsfig{file=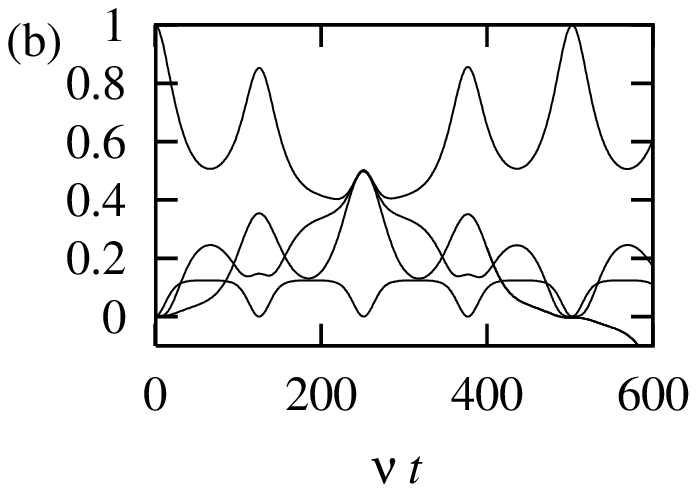,angle=270,width=5cm}
 \end{minipage}
\end{center} 
  \caption{Time evolution of density matrix elements according to
    ~(\ref{u}), with all ions initially in the ground state.  
(a) Pertubative regime (b) Fast gate. The first
    curve (counting
    from above at $\nu t\approx 1000$ in (a) and $\nu t\approx130$ in (b))
    represents $\rho_{gg,gg}$, the second is the
    imaginary part of $\rho_{gg,ee}$, the third is $\rho_{ee,ee}$, and the
    last curve is the real part of $\rho_{gg,ee}$. In (a) the physical
    parameters are $\delta=0.9\nu$, $\eta=0.1$, and $\Omega=0.1\nu$. In (b)
    the physical parameters are $\delta=0.95\nu$, $\eta=0.1$, and
    $\Omega=0.177\nu$. The parameters in (b) are chosen such that a
    maximally entangled
    state $\frac{1}{\sqrt{2}}(|gg\ra-i|ee\ra)$ is formed at the
    time $\nu t\approx250$.} 
  \label{twogates}
\end{figure}

For comparison we show in Fig.~\ref{twogates} the accomplishments
of  (a) the slow gate and (b) the fast gate evolution.
The slow gate is correctly described by Eq.~(\ref{u}), which
simplifies because $F(t)$ and $G(t)$ are always small. 
The slow gate may be stopped
when $A(t)\approx -(\Omega \eta)^2t/(\nu-\delta)$ has 
acquired its desired value, irrespective of the current, small values of 
$F(t)$ and $G(t)$. 
(For illustrational purposes, small but non-zero values of $F(t)$ and $G(t)$ 
were chosen, leading to the small fast oscillations in the figure). 
To implement the
fast gate with a specific value of $A(\tau)$, one must choose parameters
to fulfill $F(\tau)=G(\tau)=0$. In Fig.~\ref{twogates} (b)
$F(\tau)=G(\tau)=0$ is achieved at the times $\tau\nu \approx k\cdot
125$, where $k$ is any integer. 

\subsection{RISQ computing in ion traps}

\subsubsection{Feynman computing}

The trapped ions or suitable subspaces of states of the ions 
can be used to represent other physical systems with the same
Hilbert space dimension. In the spirit of Feynman's proposal
for quantum computing \cite{feynman}, the trapped ions may thus
be used for simulation of  such other systems. 

Let us consider a specific example, where
we apply Hamiltonians which are acting identically on all ions, 
{\it e.g.}, because 
laser fields extend over the whole ion cloud rather than
being focused down on one or two ions.  It follows that a 
state of the system, which is initially symmetrical under exchange of 
different ions, will  remain symmetrical. A convenient representation
of such states is given by
the eigenstates of a fictitious total angular momentum,
$|JM\rangle$, the so-called Dicke states \cite{Dicke}.
(Every single two-level ion is generically described by
$2\times 2$ Pauli spin matrices, and the associated
fictitious spin 1/2 add up to a total $J=N/2$ angular momentum.)

In the Dicke representation,
$N=2J$ is the total number of ions, and $M$ counts the
number of excited ions, $N_e=J+M$.
A single resonant laser field,
which excites all ions with same amplitude
acts as the angular momentum raising operator $J_+$ on the
symmetrical states (and the adjoint lowering operator $J_-$),
and effectively it acts as a geometrical rotation of the
state vector. Other operators like $J_y^2$ are of more interest,
in particular if the interaction can be applied to the system
in a pulsed fashion to yield  the kicked, non-linear rotor, 
which is a key example of a classically chaotic system.  

It is of course not sufficient to make the identification
between  the states of the relevant system and the states
of the trapped ions. We also have to find a way to
implement the 
collective $J_y^2$ Hamiltonian.
In terms of individual
raising and lowering operators, it is apparently necessary to introduce
interactions among all the particles of the form
of the single pair interaction described in the previous
subsection.  This, however, turns out to be easier than to 
carry out even a single two-qubit computation:
If the trap contains a larger number of 
ions, which are {\it all illuminated by the bichromatic light}, any two
ions can together resonantly perform the transitions illustrated
in Fig, 1, and the Hamiltonian automatically involves the sum over all
pairs in the trap. This sum is nothing but the collective operator 
$J_y^2$.

The interest in studying the $J_y^2$ Hamiltonian 
was recently stressed by Milburn \cite{footnote}, and it
is emphasized by Haake \cite{Haake} in this issue of J. Mod. Opt.
By a simple translation of Haake's
arguments for the atom-cavity coupling to the mathematically
equivalent trapped ion dynamics, we observe that if
ions with more than two levels are used, interaction
Hamiltonians of a more complicated structure can be tailored,
to simulate, {\it e.g.}, the 
SU(2) Lipkin model \cite{Haake}.

\subsubsection{Multiparticle entanglement}
\label{multisec}

It turns out \cite{bic2}  that the multi-atom collective operator 
$J_y^2 = (\frac{\hbar}{2}\sum \sigma_{y,i})^2$ generates a maximally
entangled state if it is applied to a whole ensemble of ground state
ions,
\begin{eqnarray}
|\Psi\rangle =|gg ... g\rangle
\rightarrow \frac{1}{\sqrt{2}}( {\rm e}^{i\phi_g}|gg ... g\rangle + 
{\rm e}^{i\phi_e}|ee ... e\rangle).
\label{ghz}
\end{eqnarray}
These states have several very interesting 
applications both in fundamental physics and technology. They are
Schr{\"o}dinger cat
superpositions of states of mesoscopic separation, and they are ideal for
spectroscopic investigations.
 
In current frequency standards the atoms or ions are independent, and
when they are interrogated by the same field, the outcome of a
measurement fluctuates as the square root of the number of atoms $N$.
The relative frequency uncertainty in samples with many atoms thus
behaves like
$\frac{1}{\sqrt{N}}$. If the duration of the measurement is shorter than
the coherence time of the atomic coherence, which is typically the case
in atomic frequency standards, by binding the ions 
together as in Eq.~(\ref{ghz}) we are sensitive to 
the Bohr frequency between $|gg ...g\rangle$ and $|ee...e\ra$ 
which is proportional
to $N$, and consequently the frequency uncertainty is proportional
to $\frac{1}{N}$ \cite{bollinger}. 
If the duration of the frequency measurement exceeds the time scale
of internal atomic decoherence $\tau_{dec}$, the shorter coherence time
$\tau_{dec}/N$ of the entangled state actually leads to the same 
resolution for that state and for an uncorrelated ensemble of atoms
\cite{Huelga}. 

The successful implementation of our proposal to produce
the state in Eq.~(\ref{ghz}) with four ions 
was  recently reported  by the NIST group in Boulder
\cite{nature}.

\section{Optical lattice quantum computers}

\subsection{General purpose quantum computing in optical lattices}

\begin{figure}
\begin{center} 
 \epsfig{file=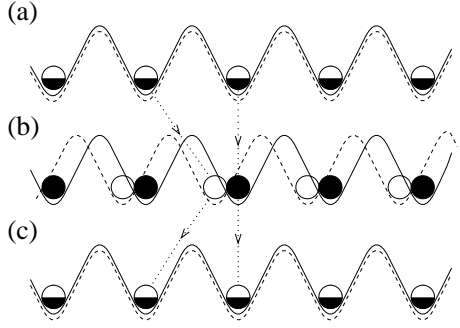,angle=270,width=6cm}
\end{center}
\vspace{0.5cm}
 \caption{(a) Two overlapping  lattices trapping the two internal states
   $|0\rangle$ (black circle) and 
   $|1\rangle$ (white circle). 
   (b) The 
   lattices are displaced  
   so that if an atom is in the $|1\rangle$ state, it
   is moved close to  the neighbouring atom if this is in
   $|0\rangle$ causing an interaction between the two atoms. (c) The
   lattices are returned to their initial position, where the
   non-interacting atoms may be driven by external fields.}
   \label{displace}
\end{figure}

In Refs. \cite{gatecirac,gatebrennen} two different methods to
perform a coherent evolution 
of the joint state of pairs of atoms  in
an optical lattice were proposed. Both methods
involve displacement of two optical lattices with respect to each
other. Each lattice traps one of the two internal states $|0\rangle$ and
$|1\rangle$ of the atoms. 
Initially, the two lattices are
on top of each other and the atoms are assumed to be cooled to the
vibrational ground state in the lattices. 
The lattice containing the $|1\rangle$ component of the wavefunction is
now displaced so that if an  atom (at the lattice site $k$) is in
$|1\rangle$, it is transferred to the vicinity of
the neighbouring atom (at the lattice site $k+1$) if this is in
$|0\rangle$, causing an interaction between the two atoms. See
Fig.~\ref{displace}.  The atoms interact through controlled collisions 
or through optically induced dipole-dipole interactions. After the interaction,
the lattices are returned to their initial position and the internal states
of each atom may be subject to single particle unitary evolution. The
displacement and the
interaction with the neighbour yields a certain
phase shift $\phi$ on the 
$|1\rangle_k |0\rangle_{k+1}$ component of the wavefunction, i.e., 
\begin{eqnarray}
  |0\rangle_k|0\rangle_{k+1}\rightarrow&
  |0\rangle_k|0\rangle_{k+1} &|0\rangle_k|1\rangle_{k+1}
  \rightarrow |0\rangle_k |1\rangle_{k+1}  
  \nonumber \\ 
  |1\rangle_k|0\rangle_{k+1}\rightarrow& e^{i\phi}|1\rangle_k
  |0\rangle_{k+1}\hspace{0.25cm}&|1\rangle_k|1\rangle_{k+1} \rightarrow
  |1\rangle_k|1\rangle_{k+1}, 
  \label{phaseshift}
\end{eqnarray}
where $|a\rangle_k$ ($a=0$ or $1$) refers to the state of the atom at the
$k$'th lattice site.
In \cite{gatecirac, gatebrennen} it is suggested to build a general purpose
quantum computer in an optical lattice based on the 
two-atom  gates in Eq. (\ref{phaseshift}) and single
atom control, which is possible by directing a laser beam on each atom. 

\subsection{RISQ computing in optical lattices}

Individual access to atoms in an optical lattice is not a 
realistic demand. The lattice sites are spaced by only a fraction of
the optical wave length, and hence focusing of a light beam
will not yield single site resolution. 
One may construct field configurations or magnetic micro-traps
with periods larger than optical wavelengths \cite{spacing} 
and still use the internal state selective 
translation and interaction 
to implement the two-qubit gate. In this section, however, we shall
show that there may be interesting possibilities in the optical
lattices, despite the lack of access to the individual atoms.
We describe how atoms in an optical lattice may be
manipulated to simulate spin-spin interactions which are used to describe
ferro-magnetism and
antiferro-magnetism 
in condensed matter physics \cite{lattice}. 
We also show that with a specific choice of
interaction we may generate spin squeezed states \cite{ueda}  which may be
used to enhance spectroscopic resolution\cite{winsq}, {\it e.g.}, in atomic
clocks.  

\subsubsection{Feynman computing in an optical lattice}

Our two level quantum systems  conveniently describe  spin $1/2$
particles with the two states $|0\rangle_k$ and $|1\rangle_k$ representing
eigenstates of the $j_{z,k}$-operator
$j_{z,k}|m\rangle_k=m|m\rangle_k,\ m_z=\pm 1/2$ ($\hbar=1$). The phase-shifted
component of the  wavefunction in
Eq.~(\ref{phaseshift}) can thus be  identified with the operator
$(j_{z,k}+1/2)(j_{z,k+1}-1/2)$, and the total evolution composed of the
lattice translations and the interaction induced phase shift 
may be described by the  unitary operator $e^{-iHt}$ with the
Hamiltonian $H=\chi(j_{z,k}+1/2)(j_{z,k+1}-1/2)$ and time $t=\phi / \chi$. In
a filled lattice all atoms are brought into contact with their
nearest neighbour according to (\ref{phaseshift}), and the
evolution is  described by the Hamiltonian $H=\chi \sum_k
(j_{z,k}+1/2)(j_{z,k+1}-1/2)$. 

If we neglect boundary terms this
Hamiltonian reduces to
\begin{equation}
 H_{zz}= \chi \sum_{<k,l>} j_{z,k}j_{z,l},
 \label{kunz}
\end{equation}
where the sum is over nearest neighbours. By appropriately displacing the
lattice  we may extend the sum to
nearest neighbours in two and three dimensions. $H_{zz}$ coincides with the
Ising-model Hamiltonian
\cite{ising,reif} introduced to describe ferro-magnetism. Hence, by elementary
lattice displacements we perform a quantum simulation of a
ferro-magnet (or of an antiferro-magnet depending on the sign of $\chi$). 
This is an extraordinary example of Feynman quantum computing
which is grossly simplified by the locality and the translational
invariance of the physical model.

A resonant $\pi/2$-pulse acting
simultaneously on all atoms rotates the $j_z$-operators into
$j_x$-operators,  $e^{ij_{y,k}\pi/2}
j_{z,k}e^{-ij_{y,k}\pi/2}=j_{x,k}$. Hence, by applying  
$\pi/2$-pulses, in conjunction with the   displacement sequence, we turn
$H_{zz}$ 
into $H_{xx}$ and $H_{yy}$, the second and third term in 
the more general Heisenberg-model Hamiltonian  \cite{Heisenberg}
\begin{equation}
 H_f=\sum_{<k,l>} \chi j_{z,k}j_{z,l}+ \eta j_{x,k}j_{x,l} + \lambda
 j_{y,k}j_{y,l}.
 \label{ferro}
\end{equation}
By adjusting the duration of the 
interaction with the neighbours we may  adjust the
coefficients $\chi$, $\eta$ and $\lambda$  
to any values. We cannot, however, produce $H_f$ by simply applying
$H_{zz}$ for the desired time $t$, followed by 
$H_{xx}$ and $H_{yy}$, because the different Hamiltonians in
Eq.~(\ref{ferro}) do not
commute. Instead we choose  short
time steps, i.e.,  small 
phase shifts $\phi$ in Eq.~(\ref{phaseshift}), 
and  by repeated application of $H_{zz}$, $H_{xx}$ and $H_{yy}$, we 
approximate the action of $H_f$ with an
error of order $\phi^2$. 

A host of magnetic phenomena may now be simulated on our optical lattice:
Spin waves, solitons,  topological excitations, two magnon bound states,
etc. 
Models for magnetic phenomena have interesting thermodynamic
behaviour and we propose to carry out calculations
for non-vanishing temperature by optically pumping a
fraction of the atoms to the $|1/2\rangle$ state. The randomness of
the pumping introduces entropy into the system and produces a
micro-canonical \cite{reif} realization of a finite temperature.
  
The results of the simulation may be read
out by optical diffraction of light, sensitive to the internal atomic
states. Although individual atoms may not be resolved, optical detection
may also be used to read out magnetic structures on a spatial
scale of a few lattice periods.

\begin{figure} 
\begin{center}
  \epsfig{file=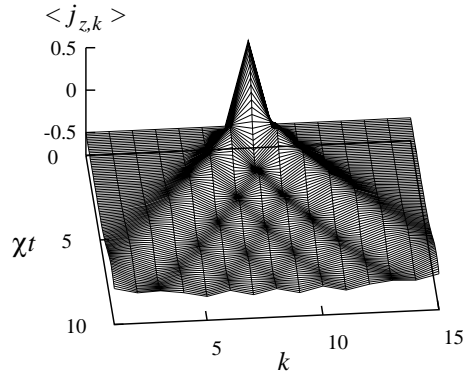,angle=270,width=6cm}
\end{center}
 \caption{Propagation of a spin wave in a one dimensional string. The
   central atom is flipped at $t=0$, and repeated application of $H_{zz}$,
   $H_{xx}$ and $H_{yy}$ results in a wave propagating to the
   left and right. The figure shows the evolution of $<j_{z,k}>$ for all
   atoms ($k$).} 
  \label{wave} 
\end{figure}

For a few atoms the system may be simulated numerically on a classical
computer. In Fig.~\ref{wave} we show the propagation of a spin wave in a
one-dimensional string of 15 atoms which are initially in the
$|-1/2\rangle$ 
state. For illustrational purposes we assume that the central spin is
flipped at $t=0$. The Hamiltonian (\ref{ferro}) which can be implemented
without access to the individual atoms then causes a spin wave to propagate to
the left and right. 

So far we have assumed that the lattice contains one atom at each
lattice site and that all atoms  are cooled to the vibrational ground
state.  The present
experimental status in optical lattices is that atoms can be cooled to the
vibrational ground-state in 2D
\cite{groundstate}. A mean filling factor of unity in 3D is reported in
\cite{depue}, but when at most a single atom is permitted at each 
lattice site  a mean occupation of 0.44 is achieved. 
The interaction in  a partially filled 
lattice may be described by the Hamiltonian
$H=\sum_{k,l} \chi_{k,l} h_k(j_{z,k}+1/2)h_l(j_{z,l}-1/2)$, where the 
stochastic variable $h_k$ is $1$ $(0)$ if a lattice site is
filled (empty), and 
where the coupling constants $\chi_{k,l}$  between atoms $k$ and $l$
vanish unless the atoms are brought into contact by the 
lattice displacements. 
If we  displace the atoms so that $\chi_{k,l}$ is symmetric in $k$ and $l$,
we produce the Hamiltonian
$ H=\sum_{k,l} \chi_{k,l} h_k j_{x,k} h_l j_{x,l}.$
This Hamiltonian models magnetism in random structures,
and it might shed light on morphology properties, and, {\it e.g.}, percolation
\cite{percolation}. 

\subsubsection{Multi-particle entanglement and spin squeezing}

Polarization rotation spectroscopy and high
precision atomic fountain clocks are
now limited by the $1/\sqrt{N}$ sensitivity discussed in Sec. \ref{multisec}
\cite{jens,precision}. In \cite{ueda} it is
suggested  to produce spin squeezed states which redistribute the uncertainty
unevenly between collective spin components like $J_x$ and $J_y$, so
that measurements, sensitive to the component with reduced uncertainty,
become more precise. Spin squeezing resulting from absorption of
non-classical light has been suggested  \cite{kuzmich} and demonstrated
experimentally \cite{jan}. Ref. \cite{ueda} presents an analysis of 
squeezing obtained from
the non-linear couplings  $H=\chi J_x^2$ and
$H=\chi(J_x^2-J_y^2)$. 
The product $J_x^2$ involves terms $j_{x,k}j_{x,l}$
for all 
atoms $k$ and $l$, and this coupling may be produced by
displacing the lattices 
several times so that the $|1/2\rangle$  component of  each atom visits
every lattice site and interacts with all other atoms.  In a
large lattice such multiple displacements  are not desirable, they may
be too difficult to control precisely, and they take too much time.
We shall show, however,  that substantial spin-squeezing occurs through
interaction with {\it only a few} nearby atoms.

If each atom visits only its nearest
neighbour, $\chi_{k,l}=\chi \delta_{k+1,l}$, we
find that the mean spin vector is in the negative $z$ direction and it has
the expectation value $<J_z>=-\frac{N}{2}\cos^2(\chi t)$.
The time dependent variance of the 
spin component $J_{\theta}=\cos(\theta)J_x+\sin(\theta)J_y$ with
$\theta=-\pi/4$  is obtained by a
lengthy, but
straightforward, calculation
\begin{equation}
 (\Delta J_{-\pi/4})^2 =\frac{N}{4}\left[1+\frac{1}{4} \sin^2(\chi
 t)-\sin(\chi t)\right]. 
 \label{dj}
\end{equation}

\begin{figure} 
\begin{center} 
 \begin{minipage}{5cm}
  \epsfig{file=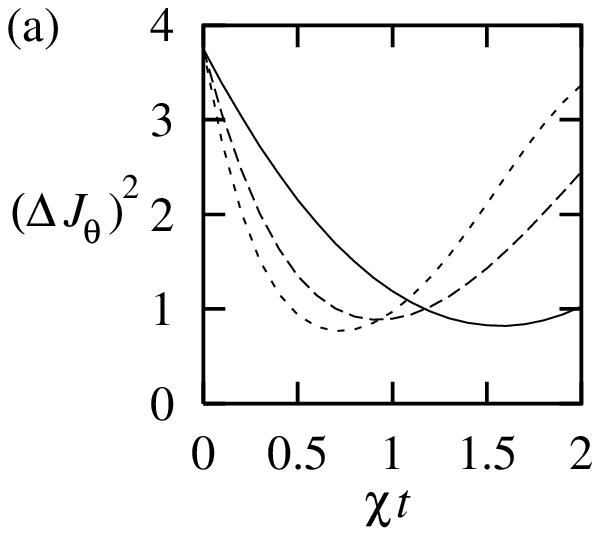,angle=270,width=5cm}
 \end{minipage}
\hspace{0.5cm}
 \begin{minipage}{5cm}
  \epsfig{file=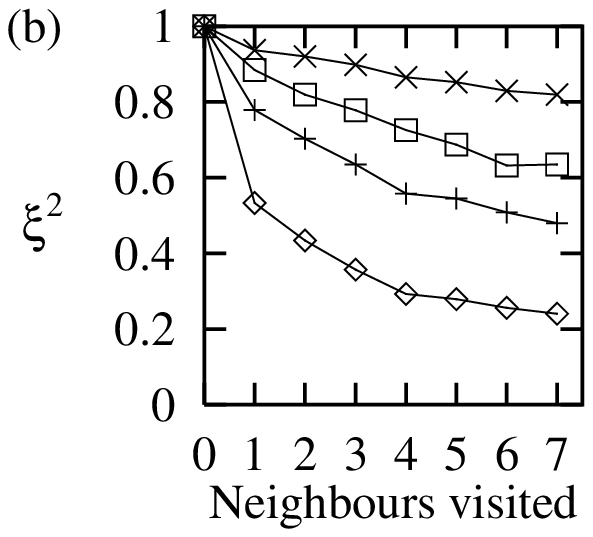,angle=270,width=5cm}
 \end{minipage}
\end{center}  
 \vspace{0.5cm}
 \caption{Squeezing in a
   one-dimensional lattice with 15 atoms. (a) Evolution of $(\Delta
   J_\theta)^2$ during interaction with 1, 2, and
   3  neighbours (full, dashed, and short dashed line, respectively). 
(b) Minimum attainable squeezing parameter $\xi^2$ for filling factors
   $p$=100\% ($\diamond$), 50\% (+), 25\% ($\Box$), and 10\% ($\times$) as
   functions of the number of sites 
   visited.  } \label{partsqueez} 
\end{figure}

Fig.~\ref{partsqueez} (a) shows the evolution of $(\Delta
J_\theta)^2$ when we visit 1, 2, and 3 neighbours.  We
assume the same phase shift for all collisions, i.e., all non-vanishing
$\chi_{k,l}$ are identical. The squeezing angle 
$\theta=-\pi/4$  is optimal 
for short times $\chi t << 1$. For longer times the optimal  angle deviates
from $-\pi/4$, and we plot the variance  $(\Delta J_{\theta})^2$ minimized
with 
respect to the angle $\theta$. If 
$\frac{1}{\sqrt{2}}
({\rm e}^{-i\theta/2}|1/2\rangle+ {\rm e}^{i\theta/2}|-1/2\rangle)$
is rotated into $|1/2\rangle$, subbinomial counting statistics of the
$|1/2\rangle$ population provides an easily accessible experimental signature
of squeezing of $J_\theta$.

In \cite{winsq}
it is shown that if spectroscopy is performed with $N$
particles, the reduction in the frequency
variance due to squeezing is given by the quantity 
\begin{equation}
  \xi^2=\frac{N\langle \Delta J_\theta\rangle^2}{\langle
    J_z \rangle^2 }, 
  \label{xi}
\end{equation}
and in
Fig.~\ref{partsqueez}  (b) we show the minimum value
of $\xi^2$ as a function of the number
of neighbours visited. 
We have performed simulations of 
squeezing in a partially filled one dimensional lattice. In our model
each lattice site contains an atom with a probability $p$, and the size
of the lattice is adjusted to accommodate 15 atoms. 
The calculations shown in Fig.~\ref{partsqueez}  (b) 
demonstrate that considerable squeezing
may be achieved by visiting just a few neighbours even in dilute lattices.

\section{Outlook}

Our examples with ion trap and optical lattice quantum
computers explicitly confirm the assumption
\cite{feynman,lloyd} that a quantum
computer aimed at the solution of a  quantum problem may be 
easier to realize in practice than a general purpose quantum computer,
because the desired solution is governed by physical interactions which are
constrained, {\it e.g.}, by locality and symmetries. 

The problems addressed here, the kicked non-linear rotor, the Lipkin
model, and the Heisenberg model of ferromagnetism and
anti-ferromagnetism, almost implement themselves in the 
quantum computer proposals with trapped ions and atoms.
The arguments are general, and one may readily conclude that
other proposals for quantum computing offer similar 
approaches to these models, and that a large variety of quantum
physics problems may be implemented much more easily than
the more mathematical algorithms of Shor \cite{shor} and Grover \cite{grover}.

If a computation can be carried  out with only few operations
and in a very short time, the problem of errors is substantially
reduced, and it seems realistic that we may soon
perform interesting calculations which are 
really impossible to carry out on a classical computer.
Extra optimism derives from the fact that small
imprecisions in our manipulation of the system translates
into small errors in the value of the physical
parameters in the simulated problem, so that, {\it e.g.}
the spin wave in Figure 4, might move at a different speed, but 
the essential physics is still preserved. The
errors are 'normal' and may well be below the precision required,
unlike the outcome of a factoring or search algorithm, where
a wrong result is useless, and
where we have to rely on the exact result to appear with finite
probability.

Both trapped ions and atoms in optical lattices
are systems which can be used in high precision
spectroscopic measurements. We have shown that
collective operators for an ensemble of ions or atoms
can be squeezed, yielding an improvement of the
precision in such measurements. 
Unlike manufactured systems, like quantum dots or
Josephson junctions, given isotopes of ions or atoms 
are identical, and they can serve as primary time standards. 
There will hence be a continuing demand to improve
experiments on these systems, irrespective of their
prospects for full scale quantum computation.
Already present atomic clocks are operating at the
projection noise limit \cite{precision}, and multi-particle 
entanglement and spin-squeezing,
in one way or another, will come in handy. 
Noise reduction derived from multi-particle
entanglement provides a macroscopic experimental signature of the microscopic 
interaction between the atoms, and hence it may help to diagnose 
gates in an atomic proto-type  quantum computer.

Quantum effects are not only subject of experimental investigation.
In the hands of experimental physicists wave mechanics is used in SQUIDs
and in atom interferometers for sensitive measurements of fields and
inertial effects; electron tunneling is used in the scanning
tunneling microscopes; the existence of discrete spectral lines 
is used for metrology; ... . Quantum information {\it is} in use
in physics, and further developments in quantum information
can find applications, ranging from the use of spin-squeezed
and Schr\"odinger cat like states to, {\it e.g.}, Grover's 
and Shor's algorithms as methods to distinguish between external
influences on a physical system \cite{farhi} and to effectively estimate 
values of complex phase factors \cite{griffith}.

We have addressed the use of quantum information as a theorist's
computational tool, and in a recent paper \cite{preskill}, Preskill
envisions the use of quantum computation for a wide
range of many-body problems. A 'symbiosis' between quantum
information and these physical problems can even be
imagined since, {\it e.g.}, topological field theories 
may in turn suggest stronger models and new algorithms for quantum computing
\cite{kitaev}.
 
Quantum computing only works, if it can be implemented on a quantum
system. There are good chances that RISQ implementations exist
for many of the physics problems amenable to quantum computing.
To identify such problems, and maybe even some mathematical
problems tractable by RISQ, is both an
interesting and useful challenge for quantum information theory.

\end{document}